\documentclass{aa}
\usepackage{aabib99}
\usepackage{graphics}
\begin{document}
\title{Analytical solution of the radiative transfer equation for polarized 
light}
\author{ A. L{\'o}pez Ariste \and M. Semel}
\institute{DASOP, URA 326, Observatoire de Paris, Section de Meudon}
\date{Received \_\_\_\_   ; Accepted  \_\_\_\_   }
\titlerunning{Analytical solution to RTE}
\thesaurus{02(02.18.7;02.16.2;)}
\offprints{A. L{\'o}pez Ariste}
\mail{Arturo.Lopez@obspm.fr}
\maketitle
\begin{abstract}
A new formalism is introduced for the transfer of polarized radiation.
Stokes parameters are shown to be four--vectors in a Minkowski-like 
space and, most strikingly, the radiative transfer equation 
(RTE) turns out to be an infinitesimal 
transformation under the Poincar{\'e} (plus dilatations) group. A solution to the 
transfer equation as a finite element of this group is proposed. 
\end{abstract}

\section{Introduction}

Since the pioneering paper of Unno \cite*{Unn56} in which for the first time
a  transfer equation was derived for polarized light in the presence of a 
magnetic field, a number of attempts have been
made to obtain a solution for it. Rachkowksy \cite*{Rac67}, after completing
the equation by adding  anomalous dispersion effects, gave the first 
analytical solution for the case of Milne-Eddington atmospheres. While 
numerical solutions were more and more successful 
\cite{Wit74,LD76,RM89,BRRCC98,art2}
analytical trials struggled to overcome the problem of non-commuting
absorption matrices. This problem, already noted in a paper of Landi
Degl'Innocenti \& Landi Degl'Innocenti \cite*{LDLD85}, is pointed out clearly
in the work of Semel \& L{\'o}pez Ariste \cite*{art1} (hereafter referred as
Paper I) as the origin of all
previous limitations. In this paper, deep insights are given on how to
benefit from the physical significance of the various terms in the transfer 
equation 
to attain an analytical solution as general as possible. Several 
transformations used
here to simplify the RTE seem to indicate that the physics of the Stokes
parameters is best described in geometrical terms.
Indeed, we shall show in this paper that the Poincar{\'e} (plus dilatations) group 
is at the origin of these geometrical aspects,
and we shall use its algebraic properties to solve the problem even for 
non--commuting  absorption matrices and give a general solution for this
equation of transfer.

The fundamental problem addressed in the present paper is the existence
of an analytical solution to the RTE. At present, analytical solutions
exist for only very limited cases and the known formal solutions \cite{LDLD85}
do not offer, in the general case, any advantage from the computational
point of view. Apart from the interest 
of a computable analytical solution by itself, it would be of greater 
importance for testing
numerical codes which nowadays are considered acceptable only by observing
their {\em convergence} upon increasing number of layers. This is not a 
very satisfactory situation. We think that the
solution presented in this paper is the first step towards a general solution
suitable for testing computations. 

Why an analytical solution could not be found so far is our first question.
It is our belief that group theory is the key to the solution and 
that explains why it had not been reached up to now (this could already be 
understood  from the conclusions in Paper I which stressed the 
importance of the non-commutativity of matrices in this problem, but it is
still clearer here). Group theory is not common in astronomical 
literature. However the RTE of polarized light has become a more and more
important  problem
in astrophysics, related, for instance, to the measurement of magnetic fields
via the Zeeman or Hanle effects. Hence group theory should be accessible
to the concerned astrophysical community, and with its help we can give
a method to find the solution, and explicitly give the full 
expression of the analytical solution for the most general case. Any other 
particular or general solution will benefit from the use
of advanced linear algebra, and so avoid wasted effort. Last but not least,
 it may deepen our understanding of Stokes polarimetry. 

We begin our research by disclosing the mathematical nature of the Stokes 
vector, starting with its physical definition and extending to the appropriate 
mathematics to treat our problem.
Extensive literature has already been devoted to the existing relations between
polarization and the Lorentz group, mainly from the optical point of 
\nocite{Clo86,GK93,SS94}
view (see for instance Cloude 1986; Givens \& Kostinski 1993; 
Sridhar \& Simon 1994 and references therein).  
Usually these works take off from the Jones formalism
for polarized light \cite{Jon41} and develop these relations. Here a 
similar path is  followed to show that the Stokes vector
 is a 4--vector in a Minkowski-like space. The demonstration is based
on the comparison between the usual definition for the Stokes parameters
\nocite{Shur62,JLS89,LL71}
(see for example Shurcliff, 1962, or Jefferies, Lites \& Skumanich, 1989) and 
the well--known relations between the definition of a spinor and its 
different representations (see for instance Landau \& Lifshitz, 1971).
In the same line of thought, we will propose in Section 3
that the RTE is just a representation  of an infinitesimal Poincar{\'e} 
(plus dilatations) transformation in the Minkowski-like space where the 
4--vectors are  best described. This new interpretation of the equation of 
transfer suggests that any solution to this equation must be a finite 
Poincar{\'e} transformation. We calculate it in Sections 4 and 5.
We note that nothing differentiates the transfer equation as written 
elsewhere in the literature from a Poincar{\'e} transformation, although at present
a demonstration is not available, apart from this similarity. We anticipate
that a finite Poincar{\'e} transformation may be a solution for the RTE.

The concepts of 4--vector, Minkowsky space or Lorentz and Poincar{\'e} 
transformations must be understood throughout this paper in their purest 
mathematical sense, beyond their historical meaning. These concepts originated in the framework of the special
theory of relativity, however mathematics did abstraction of these tools and
incorporated them into more general frames of geometry and group theory. 
Therefore we define
4--vectors as sets of 4 numbers characterized by their Minkowsky norm and 
described in a hyperbolic 4--dimensional space called the Minkowsky space,
which, in this paper, we will refer to as the {\em Minkowsky--like space}, in
order to stress the difference with the usual Minkowsky space used in 
relativity. Lorentz
and Poincar{\'e} transformations describe movements in this space: generalized
translations and rotations. In the relativistic formalism these movements are
interpreted as a change of reference system. In this paper they are not 
given  that meaning, but are seen as changes in the polarization state.
The manipulation of these concepts is identical here and in the relativistic 
formalism, if one takes care in substituting
the four Stokes parameters for space and time, and for the speed 
its analogues as shown in Section 3.

\section{ Stokes parameters as a 4--vector}

\begin{figure}[htbp]
\resizebox{\hsize}{!}{\rotatebox{-90}{\includegraphics{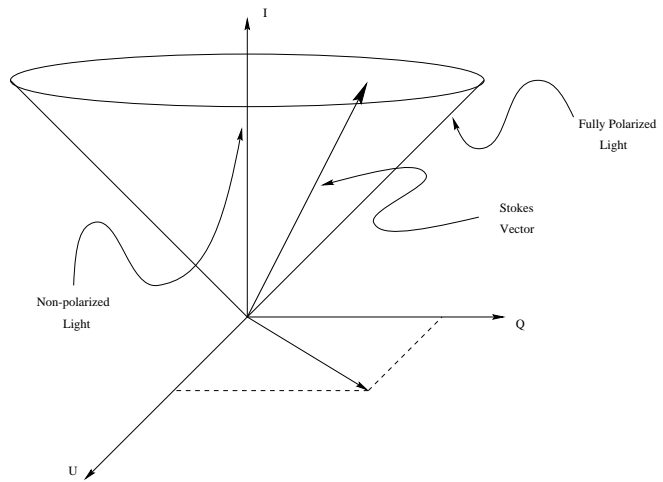}}}
\caption{3-dimensional projection of the light cone for the Stokes vector}
\label{cono}
\end{figure}

The 4 Stokes parameters are usually represented
by $I,Q,U,V,$ where  $I$ stands for the total intensity of light, 
$Q$ and $U$ for the linear polarized light in two axes rotated by $45^\circ$ 
one from the other, and $V$ for the circularly polarized light. These four 
quantities  are not completely free 
but must satisfy an energy condition: there cannot be more polarized light
than total light. This condition suggests an
 interpretation of the Stokes parameters
as a 4--vector in a Minkowski-like space. The norm of a vector in such a space
is defined as
\begin{equation}
\| \vec{I}  \| ^2 = I^2 - Q^2 - U^2 -V^2 .
\label{norm}
\end{equation}
Hence the above condition is naturally satisfied by vectors with a
positive norm, in parallelism to {\it time-like} vectors in special relativity.
Continuing this parallelism, a {\it light-cone } can be defined as the surface 
which obeys the condition
\begin{equation} \| \vec{I} \| = 0, \end{equation}
that is
\begin{equation}  I^2 = Q^2 + U^2 + V^2. \end{equation}
In our context, this surface contains all the different possibilities for 
fully polarized light. The Stokes vector must be inside  or, in the limit, on  
this {\em light-cone} (see Fig. \ref{cono}) to obey condition
 (\ref{norm}). An exception to this
parallelism with special relativity: the {\it backward light cone } does 
not have an equivalent with the  Stokes vectors.

To consolidate and extend  this interpretation of the Stokes 
parameters we  begin with the definition of the Stokes parameters in terms of 
the electric field.

A transversal monochromatic light wave  is completely 
described by 
$E_x$ and $E_y$, the components of the electrical field in a plane 
perpendicular to the direction of the propagation of light, $z$. Following  
definitions in Landau \& Lifshitz \cite*{LL71}, these two components can be 
arranged in a 2--dimensional vector.
Since it transforms linearly under the proper Lorentz group (ibid.)
 this vector can be called a {\em spinor of rank one}. As an 
illustration of  this kind of transformation, we profit from the fact 
that any element of this group in its 2--dimensional representation can 
be written as a linear combination of the Pauli matrices (plus the $2\times 2$
identity matrix): 
\begin{eqnarray}
\sigma _0 =\pmatrix {1&0\cr 0&1} ,\: \sigma _1=\pmatrix {1&0\cr 0&-1} \\
\sigma _2=\pmatrix {0&1\cr 1&0} ,\: \sigma _3=\pmatrix {0&-i\cr i&0},
\end{eqnarray}
and transform the electric field vector by these matrices:
\begin{eqnarray}
{\bf \sigma }_0 \pmatrix{E_x \cr E_y } =  \pmatrix{E_x \cr E_y}, &\:
{\bf \sigma }_1 \pmatrix{E_x \cr E_y} =  \pmatrix{E_x \cr -E_y} \nonumber \\
{\bf \sigma }_2 \pmatrix{E_x \cr E_y} =  \pmatrix{E_y\cr E_x}, &\:
{\bf \sigma }_3 \pmatrix{E_x \cr E_y} =  \pmatrix{-i E_y \cr i E_x}.
\end{eqnarray}
The result is always a new 2-dimensional electric field vector which describes
a different state of polarization.

A spinor of rank two can be easily constructed by multiplying conveniently 
two spinors of rank one. For our particular spinor we obtain 
$$ \tens{J}'=\pmatrix { E_xE_x^* & E_xE_y^* \cr E_yE_x^* & E_yE_y^* } ,$$
where the $*$ symbol stands for complex conjugated.
This 2nd rank spinor is to be compared with the {\it coherency matrix} (see 
Born \& Wolf 1980). In fact the last is defined for 
any given light beam, and a  \nocite{BW80}
mean over frequencies or time is necessary in the above expression of 
$ \tens{J}'$ to fulfill the definition. The average of $\tens{J}'$
is a linear combination of matrices of the form referred. 
Since spinors of rank two form a linear vector space, any linear combination 
of spinors will be a spinor as well. Hence the coherency matrix, defined 
usually as
$$ \tens{J}=\pmatrix {<E_xE_x^*> & <E_xE_y^*> \cr <E_yE_x^*> & <E_yE_y^*> }, $$
is still a spinor of rank two. 

Since a spinor built in this way has 4 independent 
components (the four entries of the matrix), there must exist
a relation between it and a 4-vector, which also has 4 independent components. 
Technically speaking, both must be  different 
{\em realizations of the same irreducible representation of the Lorentz group}
(Landau \& Lifshitz 1971, page 55).
The components $(I,Q,U,V)$ of this 4--vector are indeed related to the
components of $\tens{J}$ as
\begin{eqnarray}
I &=& \frac{1}{2}(\tens{J}_{11} + \tens{J}_{22}), \nonumber \\
Q &=& \frac{1}{2}(\tens{J}_{11} - \tens{J}_{22}), \nonumber \\
U &=& \frac{1}{2}(\tens{J}_{12} + \tens{J}_{21}), \nonumber \\
V &=& \frac{1}{2}i (\tens{J}_{12}- \tens{J}_{21}). 
\label{IJ}  \end{eqnarray}
These components are identical to the definition of the Stokes parameters
as given by  Jefferies {\it et. al.} \cite*{JLS89}, or Born \& Wolf 
\cite*{BW80} for instance.
The conclusion is evident: the resulting 4--vector, derived from
the coherency matrix, a spinor of rank 2, is the Stokes vector.
This formalism can be given in an alternative way: an usual basis for spinors
is the set of Pauli matrices plus the $2\times 2$ identity matrix.
The coherency matrix expressed in this basis has for coefficients the Stokes 
parameters
\begin{equation}
\tens{J} = I \sigma _0 + Q \sigma _1 + U\sigma _2 + V\sigma _3 \end{equation}
Using this relation we can write in a more compact form relations  (\ref{IJ})
as:
\begin{equation}
\vec{ I} = \frac{1}{2} Tr \left[ \tens{J}{\bf \sigma} \right], \end{equation}
where the vector $ {\bf \sigma}$ has four components, the $2\times 2$ identity 
matrix and the 
3 Pauli matrices, and {\it Tr } denotes the trace operation on  matrices.

These relations stress further  the  interpretation of the Stokes parameters 
as a 
4--vector  in a Minkowski-like space. We stress that  when talking about a 
Minkowski--{\em like} space we mean that
the coordinates in our 4--dimensional space are no longer space and time, but
the Stokes parameters, contrary to the usual Minkowski
space used in relativistic formalism. On the other hand the underlying 
{\em hyperbolic} geometry is exactly the same in both cases, mathematically 
speaking  they are the same space. All the usual properties of the 
Stokes parameters are 
recovered in this space. For instance the sum of two 4--vectors is a new 
4--vector, a well-known property of the Stokes parameters. Contrary to the 
usual Minkowsky space, the absence of the
{\em backward light cone} implies not only that negative intensities are 
meaningless, but also that the negative of a Stokes vector does not exist. 
Hence subtraction of Stokes vectors is naturally forbidden in this space.

\section{Radiation transfer equation as a Poincar{\'e} (plus dilatations) 
infinitesimal  transformation}
\label{lorentz}
Since the Stokes vector is a 4--vector in a Minkowski-like space, one may
wonder what would be the 
meaning of a Lorentz transformation over the Stokes parameters. 

Homogeneous Lorentz transformations form a 6--parameter Lie group: 
6 generators suffice to describe all possible infinitesimal
\nocite{Grei90}
transformations. These generators are (see for example Greiner 1990, or any
textbook in special relativity or group theory)\footnote{Everywhere in this
paper we use $g=diag(1,-1,-1,-1)$ as the metric for the Minkowski space}:
\begin{itemize}
\item The three  $4\times 4$ matrices for  3-dimensional spatial rotations
$$ \tens{S}_Q=\pmatrix{0&0&0&0\cr 0&0&0&0\cr 0&0&0&1\cr 0&0&-1&0\cr} ,
   \tens{S}_U=\pmatrix{0&0&0&0\cr 0&0&0&-1\cr 0&0&0&0\cr 0&1&0&0\cr} ,$$
$$   \tens{S}_V=\pmatrix{0&0&0&0\cr 0&0&1&0\cr 0&-1&0&0\cr 0&0&0&0\cr} $$

\item The three $4\times 4$  matrices for  hyperbolic rotations (or
Lorentz boosts in relativistic terms) 
$$ \tens{K}_Q=\pmatrix{0&1&0&0\cr 1&0&0&0\cr 0&0&0&0\cr 0&0&0&0\cr} ,
   \tens{K}_U=\pmatrix{0&0&1&0\cr 0&0&0&0\cr 1&0&0&0\cr 0& 0&0&0\cr} ,$$
$$   \tens{K}_V=\pmatrix{0&0&0&1\cr 0&0& 0&0\cr 0&0&0&0\cr 1&0&0&0\cr} $$
\end{itemize}
An infinitesimal Lorentz transformation over the Stokes 4--vector $\vec{I}$ 
can be expressed as a sum
of generators multiplied by their respective infinitesimal parameters: 
\begin{equation}
\vec{I}' = \vec{I} 
+\sum_{i=Q,U,V} \beta_i \tens{S}_i
\vec{I} +\sum_{i=Q,U,V} \gamma_i \tens{K}_i \vec{I} .\end{equation}
For convenience, we can re-write all these parameters($\beta_i,\gamma_i$) in
terms of a  common parameter $d\tau $, expressed in differential form (we are 
dealing with an infinitesimal transformation)
\begin{eqnarray}
\vec{\beta} & = & (\beta_Q , \beta_U,\beta_V)=- (\rho _Q,\rho _U,\rho _V)\cdot 
d\tau ,\\
\vec{\gamma} & = & (\gamma_Q ,\gamma_U,\gamma_V)=- (\eta _Q,\eta _U,\eta _V)\cdot  d\tau .   
\end{eqnarray}
And the infinitesimal transformation reads now:
\begin{equation}
\vec{I}' = \vec{I} - \left( 
\sum_{i=Q,U,V} \rho _i \tens{S}_i + 
\sum_{i=Q,U,V} \eta _i \tens{K}_i \right) \vec{I} d\tau .\end{equation}
Given the infinitesimal character of the transformation, expressed explicitly
by putting the common parameter $d\tau$, 
it easily leads to a differential
equation for $\vec{I}$ in the variable $\tau $:
\begin{equation}
\frac {d}{d\tau } \vec{I} = -\tens{K'} \vec{I} ,
\end{equation}
where $\tens{ K'}$ is a $4\times 4$ matrix given by
$$
\tens{ K'}=\sum_{i=Q,U,V} \rho _i \tens{S}_i + \sum_{i=Q,U,V}
\eta _i \tens{K}_i =$$
$$=\pmatrix{ 0 & \eta _Q & \eta _U & \eta _V \cr
          \eta _Q & 0 & \rho _V &-\rho _U \cr
	\eta _U &-\rho _V &0  &\rho _Q \cr
	\eta _V&\rho _U&-\rho _Q& 0 } .$$
Matrix $\tens{ K'}$ looks like the well-known absorption matrix \cite{LD92}.
At this point the possibility that the RTE could be written as an infinitesimal
transformation involving the Lorentz group seems to be at hand. But  
$\tens{ K'}$
still differs from the general form of the absorption matrix. In particular the
inclusion of a diagonal term is necessary if we want to take into account
the scalar absorption represented by $\eta _I$ in the usual matrix.
To include it we just add to the usual 6 generators of the Lorentz group the
one for the dilatation transformation \footnote{The dilatation transformation 
does not belong to the usual Lorentz group. A usual definition of the 
transformations belonging to this group is that they do not change the 
Lorentz norm of any 4--vector. By definition, a dilatation transformation does 
change this norm. Fortunately the new set of 
7 generators is still  a group.}.
While dealing with the homogeneous group, we can represent the generator 
of dilatations by
the  $4\times 4$ identity matrix. We will repeat all
the previous steps calling the new parameter for this transformation 
$\eta _I$:
$$\vec{I}' = \vec{I} - \left( \eta _I \bbbone + 
\sum_{i=Q,U,V} \rho _i \tens{S}_i + 
\sum_{i=Q,U,V} \eta _i \tens{K}_i \right) \vec{I} d\tau ,$$
to finally obtain:
\begin{equation}
\frac {d}{d\tau } \vec{I} = -\tens{K} \vec{I} ,
\label {ETR}
\end{equation}
where $\tens{ K}$ is given by
$$
\tens{ K}=\eta_I \bbbone + \sum_{i=Q,U,V} \rho _i \tens{S}_i + \sum_{i=Q,U,V}
\eta _i \tens{K}_i =$$
$$=\pmatrix{ \eta _I & \eta _Q & \eta _U & \eta _V \cr
          \eta _Q &\eta _I & \rho _V &-\rho _U \cr
	\eta _U &-\rho _V &\eta _I &\rho _Q \cr
	\eta _V&\rho _U&-\rho _Q&\eta _I} .$$
Matrix \tens{K}  can now be compared to the so-called absorption matrix 
which  appears in RTE \cite{LD92,JLS89}.
Coefficient  $\eta _I$ gives the scalar  absorption, 
independent of 
polarization. This absorption has its equivalent in a  contraction (that is 
a negative dilatation) of the Stokes 4--vector. 
The 3--vector $\vec{\eta}$ is responsible for
the creation and absorption of polarization, which is understood here to be
a  hyperbolic rotation (or Lorentz boost) of  the Stokes 4-vector. Finally  
the 3--vector $\vec{\rho}$ gives the so-called Faraday rotation in Zeeman 
effect and, as its
name seems to indicate, it rotates the Stokes 4--vector inside the 
3-dimensional space of polarized states (also called 
{\em Poincar{\'e} sphere}). 
Note that from the previous paragraph it cannot be stated that any 
infinitesimal Lorentz (plus dilatations) transformation is a
transfer equation. We  state that the
absorption matrix as it is usually written elsewhere in the literature,
 cannot be
differentiated from an infinitesimal Lorentz transformation (indeed we have
decomposed the absorption matrix in the RTE in terms of the infinitesimal
generators of the Lorentz transformations). 
In fact, it is well known that the actual absorption matrix for Zeeman effect
must still obey some further constraints. In this sense a general Lorentz (plus
dilatations) transformation is too general: absorption matrices standing for
a real physical process 
would constitute only a subset of all possible Lorentz (plus dilatations)
transformations. While it is evident that a further study of this 
relation is necessary, in this paper  we will only make use of the 
{\em mathematical advantage} and postpone the rest for a forthcoming paper.

Eq. (\ref{ETR}) is not yet the complete transfer equation. An inhomogeneous 
term, the emission vector, is still needed. For this purpose there are 
appropriate movements  in the Minkowski space: the homogeneous Lorentz group 
can be extended to the inhomogeneous Poincar{\'e} group. This 10-parameter Lie 
group shares 6 infinitesimal
generators with the Lorentz group and adds 4 more generators ($\tens{P}_I, 
\tens{P}_Q, \tens{P}_U$ and $ \tens{P}_V$) to take into account 
 translations along I, Q, U and  V (what in relativistic formalism would be 
translations in time and space). These generators operate on a generic Stokes
vector as follows:
\begin{eqnarray}
\tens{P}_I \vec{I}=\pmatrix{1\cr 0\cr 0\cr 0},
\tens{P}_Q \vec{I}=\pmatrix{0\cr 1\cr 0\cr 0}, \nonumber \\
\tens{P}_U \vec{I}=\pmatrix{0\cr 0\cr 1\cr 0},
\tens{P}_V \vec{I}=\pmatrix{0\cr 0\cr 0\cr 1}.
\label{lasP}
\end{eqnarray}

After including dilatations, the infinitesimal inhomogeneous transformation 
is given by 
\begin{equation}
\vec{I}'=\vec{I} - \tens{K} \vec{I} d\tau + d\tau \sum _{i=I,Q,U,V} j_i 
\tens{P}_i\vec{I} =\vec{I} - \tens{K} \vec{I} d\tau + \vec{J} d\tau \end{equation}
where $\vec{J}$ is the emission vector. 
Some algebraic manipulations equivalent to those used for the homogeneous 
group will lead to the complete radiative transfer equation.

It is interesting to note the fact that these 
generators allow one to write an inhomogeneous term (the source function) in a
pseudo-homogeneous way:
$$ \vec{I}'=\vec{I} + \left( -\tens{K} + \sum _{i=I,Q,U,V} j_i  \tens{P}_i\right) \vec{I} d\tau.$$
A useful representation to understand this apparent paradox is the one using 
differential operators. Let us call
$\vec{I}_I = I$, $\vec{I}_Q = Q$ and so on, and write 
$$  \tens{P}_i = \frac{\partial}{\partial \vec{I}_i}$$
where $i=I,Q,U,V$. It is evident that with such an operator, relations 
(\ref{lasP}) hold. All the other generators can be rewritten in this 
representation. For instance, the dilatation generator can be written
$$ \tens{D} = \sum _{i=I,Q,U,V} \vec{I}_i \frac{\partial}{\partial \vec{I}_i} $$
and the generator of rotations in the plane $UV$, for instance, is
$$ \tens{S}_Q = -\vec{I}_U \frac{\partial}{\partial \vec{I}_V}+  \vec{I}_V \frac{\partial}{\partial \vec{I}_U},$$ 
The interested reader will find
 good discussions on the representations of the Poincar{\'e} group and the 
dilatation transformation  in Greiner (1990), Gourdin (1982) or Jones (1996) 
for example. \nocite{Jon96}
\nocite{Gou82}

\section{Finite transformations as a solution}

In view of the results obtained in the last section, the solution to the RTE 
appears to be quite straightforwardly a finite Poincar{\'e} (plus dilatations) 
transformation.
The important fact now is that we already know how to write such a finite 
transformation: if we denote the 11 generators
by $\tens{t}_i$, a finite element of this group can always be
written as
$$ \exp \left( \sum _i \xi _i \tens{t}_i\right) ,$$
where $ \xi _i$ are the  parameters of the transformation for each 
movement, the equivalent of angles for usual rotations. The next problem
is how to calculate those finite parameters  $ \xi ^i$ from their infinitesimal
counterparts ( $\eta _I , \eta _Q$ and so on). Unfortunately this is not an 
easy task. The problem resides in the non--commutativity of the generators. 
The next paragraph proposes a solution to this problem, but first we 
consider it 
useful to clarify why other approaches will not work.
Magnus \cite*{Mag54} (a brief introduction to this paper can be found
in the Appendix of the paper by Semel \& L{\'o}pez Ariste, 1999) has already 
written 
a finite transformation in terms of such a unique exponential. A quick 
inspection  of the
expression given there illustrates why we consider that this calculation is 
not an
easy task. Nevertheless some simple cases can be proposed in which the
relation between the $\xi ^i$ and the respective infinitesimal parameters is
plain. For instance in the case of 3--dimensional rotations it is always 
possible to transform (by means of the Euler angles) our initial reference 
system into another one for which the rotation axis is parallel to one of the 
new reference axes. In the case of a fixed axis, just one generator suffices 
to describe the movement.
If we call this generator $\tens{t}$, and the infinitesimal parameter $d\xi$, 
the finite transformation results in
$$ \exp \left( \xi  \tens{t}\right)=\exp \left( \tens{t} \int _C d\xi  \right).
$$
Apparently, our problem is solved if we deal with a unique generator per
exponential. Hence, we can propose for instance a solution in the form of a 
product of several 
exponentials, one for each generator:
$$\prod _{i=1,11} \exp \left(\xi _i \tens{t}_i\right) .$$
Now the derivative of each exponential in the product can be easily 
calculated in a very compact form, and so the product of all of them, and
consequently the  new finite parameters $\xi _i$ (as in the above example). 
But a new question arises: 
which order should be chosen for the exponentials? Again
due to the non-commutativity of the infinitesimal generators, 
different orders produce different results,
 \begin{equation}
\mbox{e}^{\xi _i \tens{t}_i}\mbox{e}^{\xi _j \tens{t}_j}\neq 
\mbox{e}^{\xi _j \tens{t}_j}\mbox{e}^{\xi _i \tens{t}_i} ,
\end{equation}
and in general no particular order will be the solution. 
An answer to this kind of problem has been given by Wei \& Norman
\cite*{WN63}.  In what follows, we  
sketch the solution there proposed and apply it to our particular problem. 
We start with the homogeneous equation (just  
the Lorentz group plus dilatations) and in the next section we will 
incorporate the  inhomogeneous part and handle the full Poincar{\'e} group. 

The ordering problem can be recast as follows: we may say that by introducing
an order in the exponentials we introduce an error. 
In spite of this error, let us choose a particular order for the exponentials 
and then substitute the $\xi _i$ (which could be calculated straightforwardly
as the integrals over the path of the corresponding infinitesimal parameters)
for some unspecified scalar functions $g _i$:
 \begin{equation}
\prod _{i=1,7} \exp \left(g _i \tens{t}_i\right) .$$
\label{SI}
\end{equation}
The new functions  $g _i$ have to, in a certain sense, take into account the 
effect of the $\xi _i$ and correct the error introduced by
 the chosen order. Existence for those $g_i$ functions can only be  ensured 
after introducing the proposed solution into the RTE. The following consistency
equation for the  $g_i$'s is obtained
as a necessary condition for Eq.(\ref{SI}) to be a solution:
 \begin{equation}
-\tens{K}(\tau)=\sum_{i=1,7} \dot{g}_i(\tau)\left[ \prod_{j=1}^{i-1} 
\exp (g_j \tens{t}_j )\right] \tens{t}_i\left[ \prod^{1}_{j=i-1}
\exp (-g_j \tens{t}_j) \right] ,
\label{WN}
\end{equation}
where the dot denotes derivative over the integration variable $\tau$.
This equation was the aim of this section so far. A less heuristic
but more direct way of introducing it is to look for a solution of the form
(\ref{SI}) and introduce it into the transfer equation. It is straightforward 
to fall upon Eq.(\ref{WN}) as the condition for (\ref{SI}) to be a solution.

In what follows in this and the next sections we shall solve Eq.(\ref{WN}) for
the $g_i$'s. The details being quite technical, the reader may 
wish to skip this and go directly to Eq. (\ref{total}).

Eq. (\ref{WN}) is quite an involved equation. It  requires calculation 
of a non negligible number of expressions of the form
$$\exp (g_j \tens{t}_j) \tens{t}_i \exp (-g_j \tens{t}_j).$$ This is 
to be done by means  of the Baker-Hausdorff formula, which states that 
$$ \mbox{e}^X Y \mbox{e}^{-X}=Y+[X,Y]+\frac{1}{2!}[X,[X,Y]]+
\frac{1}{3!}[X,[X,[X,Y]]]+\ldots $$

In the form they have been written in the last section, the 
generators of the Lorentz group obey the following Lie algebra:
\begin{eqnarray*}
\left[ \tens{K}_i,\tens{K}_j \right] &=&- \varepsilon _{ijk} \tens{S}_k  \\
\left[ \tens{S}_i , \tens{S} _j \right]& =& \varepsilon _{ijk} \tens{S}_k \\
\left[\tens{S}_i,\tens{K}_j\right]&=& \varepsilon _{ijk} \tens{K}_k \\
\left[\tens{S}_i,\bbbone \right]&=&0 \\
\left[\tens{K}_i,\bbbone \right]&=&0,
\end{eqnarray*}
which does not facilitate  calculations at all. A suitable combination
 of generators will yield a new base of
generators with a gentler (from our point of view) Lie algebra. 
For instance, the following set, 
\begin{eqnarray*}
\tens{H}_1&=&\tens{S}_Q + i \tens{K}_Q ,\\
\tens{H}_2&=&( \tens{K}_U-\tens{S}_V ) - i ( \tens{K}_V+\tens{S}_U ) ,\\
\tens{H}_3&=&( \tens{K}_U+\tens{S}_V ) + i ( \tens{K}_V-\tens{S}_U ) ,\\
\tens{L}_i &=& \tens{H}_i^* \mbox { for  } i=1,2,3 , 
\end{eqnarray*}
to which we add the identity, $\bbbone ,$ for completion, obeys the following 
commutation rules
\begin{eqnarray*}
\left[\tens{H}_1,\tens{H}_2\right]&=&2i \tens{H}_2 ,\\
\left[\tens{H}_1,\tens{H}_3\right]&=&-2i \tens{H}_3 ,\\
\left[\tens{H}_2, \tens{H}_3\right]&=&-4i \tens{H}_1 ,\\
\left[\tens{L}_1,\tens{L}_2\right]&=&-2i \tens{L}_2 ,\\
\left[\tens{L}_1,\tens{L}_3\right]&=&2i \tens{L}_3 ,\\
\left[\tens{L}_2, \tens{L}_3\right]&=&4i \tens{L}_1 ,\\
\left[\tens{H}_i,\tens{L}_j\right] &=& 0  , \forall i,j .
\end{eqnarray*}
The initial algebra of 6 generators has been decomposed into
two sub-algebras of 3 generators each, with the particularity
that each generator of one sub-algebra commutes with every generator of the
other one. The dilatations generator, which must be added to them, commutes 
with every other generator (remember that, while constrained to the
homogeneous group, the dilatation generator can be represented by the identity
matrix) and therefore there is no ordering problem associated with it. The 
initial problem of ordering 7 exponentials is reduced to 
ordering a subset of 3 exponentials; the order of each one of the $\tens{H}$'s with respect to the  $\tens{L}$'s generators or  the  identity being 
immaterial. 

We can rewrite solution (\ref{SI}) using the new set of infinitesimal 
generators  in explicit form as
\begin{equation}
\vec{I} (\tau) = \mbox{e}^{g_3(\tau ) \tens{H}_3}\mbox{e}^{g_2(\tau ) \tens{H}_2}
\mbox{e}^{g_1(\tau ) \tens{H}_1}\mbox{e}^{g_6 (\tau )\tens{L}_3}\mbox{e}^{g_5 (\tau )\tens{L}_2}
\mbox{e}^{g_4 (\tau )\tens{L}_1}\mbox{e}^{g_7 (\tau )\bbbone} \vec{I} (\tau _0), 
\label{SFI}
\end{equation}
where a very special order has already been chosen. Different orders, while
yielding equivalent solutions, can make calculations affordable or desperate.
With this problem in mind we have chosen a particular order.

We now express 
$\tens{K}$ in the new basis:
$$\tens{K}=\sum _i a_i \tens{H}_i + \sum _i b_i \tens{L}_i +\eta _I \bbbone $$
where
\begin{eqnarray*}
a_1&=&-\frac{1}{2}(\rho _Q + i \eta _Q) \\
a_2&=&\frac{1}{4}[ (\eta _U+\rho _V)+i(\eta _V-\rho _U)] \\
a_3&=&\frac{1}{4}[ (\eta _U-\rho _V)-i(\eta _V+\rho _U)] \\
b_i&=&a_i^* \mbox{     }, \forall i=1,2,3.
\end{eqnarray*}
Calculation of the Baker--Hausdorff series for the new generators simplifies a 
lot, and Eq. (\ref{WN}) reads now:
\begin{eqnarray}
-\sum _i a_i \tens{H}_i - \sum _i b_i \tens{L}_i - \eta _I \bbbone = 
\dot{g}_7\bbbone+\dot{g}_3\tens{H}_3 + \nonumber \\ +\dot{g}_2\left[ \tens{H}_2 +
4ig_3\tens{H}_1-4g_3^2 \tens{H}_3\right] + \nonumber \\  
+  \dot{g}_1\left[-2ig_2\tens{H}_2+\left(1+8g_2g_3\right)
\tens{H}_1 + \left(2ig_3+8ig_3^2g_2\right)\tens{H}_3\right] + \nonumber \\ 
+ \dot{g}_6\tens{L}_3 +  \dot{g}_5\left[ \tens{L}_2 - 4ig_6\tens{L}_1-4g_6^2
\tens{L}_3\right] +
\nonumber \\  
 + \dot{g}_4\left[2ig_5\tens{L}_2+\left(1+8g_5g_6\right)
\tens{L}_1-\left(2ig_6+8ig_6^2g_5\right)\tens{L}_3\right].
\end{eqnarray}
Comparing coefficients on both sides, and after proper rearrangement the 
following  set of differential equations is obtained:
\begin{eqnarray*}
\dot{g}_1 &=& -a_1 + 4i g_3 a_2 \\
\dot{g}_2 &=&-2ig_2 a_1 - (1+8g_2g_3) a_2 \\
\dot{g}_3 &=&  2ig_3 a_1 + 4g_3^2 a_2 - a_3 \\
\dot{g}_4 &=& -b_1 - 4ig_6 b_2 \\
\dot{g}_5 &=&   2ig_5 b_1 - (1+8g_5g_6) b_2 \\
\dot{g}_6 &=&-2ig_6 b_1 + 4g_6^2 b_2 - b_3 \\
\dot{g}_7 &=&-\eta _I 
\end{eqnarray*}
All of them share the same boundary condition, namely $g_i (\tau _0) = 0$, to 
satisfy the boundary condition of the RTE. 

The function $g_7$ can be integrated at once to give
$$ g_7(\tau )=-\int _{\tau _0}^{\tau } \eta _I (\tau ' ) d\tau '. $$
As expected, because of
the particularities of the Lie algebra, the set of equations for $g_{1,2,3}$ is
separated from the one for $ g_{4,5,6}$, and each  set
is the complex conjugated of the other, so that  the solution to  
$ g_{4,5,6}$ is  straightforward once the one for $g_{1,2,3}$ is given. 
Furthermore each set can be solved by quadrature, as equations for $g_1$
 and $g_2$ depend only on $g_3$, whose equation is disentangled from the 
others:
\begin{equation}
\dot{g}_3 =  \alpha g_3^2 +\beta g_3   + \gamma ,
\label{ricatti}
\end{equation}
where we have defined $ \alpha =4a_2$, $\beta= 2ia_1$ and $\gamma = -a_3$.
This is a Riccati equation, and for its solution the explicit dependences of
$a_{1,2,3}$ on the integration variable $\tau$ are required. For a constant 
$\tens{K}$ matrix the solution is straightforward, and from it those of
$g_2$ and $g_1$. More complex dependences must be carefully managed (see
\nocite{CR98}
for example Cari{\~n}ena \& Ramos \cite*{CR98} and references 
therein  for  the integrability conditions of the Riccati equation).

Before passing to the next section, where we will generalize the method
to the full Poincar{\'e} group, we go back to Eq.(\ref{SFI}). Once we have 
integrated the Riccati equation and obtained all the $g_i$'s, we still need 
to calculate the exponentials. To this end we profit from a remarkable 
property of matrices $ \tens{H}_i$ and $\tens{L}_i$:
\begin{eqnarray*}
\tens{H}_1^2=\tens{L}_1^2= -\bbbone \\
\tens{H}_2^2=\tens{H}_3^2=\tens{L}_2^2=\tens{L}_3^2=0 ,
\end{eqnarray*}
by means of which: 
\begin{eqnarray*}
\mbox{e}^{g_1 \tens{H}_1} &=& \cos g_1 \bbbone +\sin g_1 \tens{H_1} \\
\mbox{e}^{g_2 \tens{H}_2} &=& \bbbone + g_2 \tens{H}_2 \\
\mbox{e}^{g_3 \tens{H}_3} &=& \bbbone + g_3 \tens{H}_3 \\
\mbox{e}^{g_4 \tens{L}_1} &=& \cos g_4 \bbbone + \sin g_4 \tens{L_1} \\
\mbox{e}^{g_5 \tens{L}_2} &=& \bbbone + g_5 \tens{L}_2 \\
\mbox{e}^{g_6 \tens{L}_3} &=& \bbbone + g_6 \tens{L}_3 .
\end{eqnarray*}
The final complete solution for the homogeneous part  results in 
\begin{eqnarray}
\vec{I} (\tau )&=&\left[ \bbbone + g_3(\tau ) \tens{H}_3 \right] \cdot 
\left[ \bbbone + g_2(\tau ) \tens{H}_2 \right]\nonumber \\&&
\left[ \cos g_1(\tau ) \bbbone + \sin g_1(\tau ) \tens{H_1}\right ]\nonumber \\
 &&\cdot\left[ \bbbone + g_6(\tau ) \tens{L}_3 \right]\cdot  
\left[ \bbbone + g_5(\tau ) \tens{L}_2 \right] \cdot \nonumber \\
&& \left[ \cos g_4 (\tau ) \bbbone + \sin g_4(\tau ) \tens{L_1} \right ]
\cdot \exp [g_7(\tau ) \bbbone ] \vec{I} (\tau _0) .
\label{homosol}
\end{eqnarray}
The validity of this solution is almost evident: Its derivative results in 
the RTE just by making use of the differential equations satisfied by the
functions $g_i$.

This is a solution to the homogeneous equation
$$\frac{d}{d\tau}\vec{I}_L=-\tens{K}\vec{I}_L.$$
Let us write this solution as
$$\vec{I}_L (\tau)= \tens{O} (\tau,\tau_0) \vec{I}_L (\tau _0),$$
where the explicit form of $\tens{O} (\tau,\tau_0)$ can be found by comparing 
this
expression with the complete one in Eq.(\ref{homosol}). This operator 
$\tens{O} (\tau,\tau_0)$ is often referred to as  the 
{\em evolution operator} (see mainly 
Landi  Degl'Innocenti \& Landi  Degl'Innocenti 1985, who first introduced it). 
\nocite{LDLD81}
This operator trivially obeys the homogeneous equation
\begin{equation}
\frac{d}{d\tau}\tens{O} (\tau,\tau_0)=-\tens{K}\tens{O} (\tau,\tau_0),
\end{equation}
with initial condition
$$\tens{O} (\tau _0,\tau_0)= \bbbone .$$ 
Eq. (\ref{homosol}) provides on its own a general analytical solution for 
the evolution operator. 

This solution is a fully general expression for finite Lorentz 
transformations plus dilatations. But
radiative transfer cases do not cover the full spectrum of Lorentz 
transformations. In this sense the obtained solution is too general, in 
agreement with Section 3. As an 
illustration, consider the case when the $\rho$'s and $\eta$'s
are zero except for $\eta _Q$. The function $g_1$ becomes 
$$g_1=-\frac{i}{2}\int \eta _Q d\tau ,$$
and a term of the form 
$$  \cos g_1 = \cosh \frac{1}{2}\int \eta _Q d\tau $$ 
appears in the final solution. The hyperbolic cosine grows monotonously with
its argument, therefore the intensity of the
out-coming light would grow also monotonously for a semi-infinite atmosphere. 
This is a completely nonsensical result.
To recover physical sense one must impose some constraint on the allowed 
transformations. 
This constraint evidently imposes a relation between $\eta _Q$ and $\eta_I$,
whose explicit form is outside the scope of this paper, but which should be
derived from the assumed physical processes.
This example can be extrapolated to all the $\eta$'s and $\rho$'s. The 
relations thus obtained will constrain the Lorentz transformations to a
subset of matrices for which, nevertheless, the above 
solution (\ref{homosol}) will remain valid. 

\section{ Solution for the complete inhomogeneous equation}
\label{homo}
To solve the inhomogeneous equation, one would need to repeat the calculations
shown  in
the previous section, but this time for the whole Poincar{\'e} group. To 
recalculate
everything with 4 more generators involves a lot of work. The paper 
by  Wei \& Norman \cite*{WN63} provides us with a way to avoid some of this 
work. The Poincar{\'e} group  can be decomposed into the direct sum of a 
semi-simple algebra $L$ and a radical $R$ (whose definitions 
 can be found in that same paper for instance).
In terms of the previously used generators of the Poincar{\'e}
group, the semi-simple algebra is given by
$$ L=\{ \tens{H}_1,\tens{H}_2,\tens{H}_3,\tens{L}_1,\tens{L}_2,\tens{L}_3\} ,$$
the generators of the homogeneous part. The radical
is given by
$$ R=\{ \bbbone , \tens{P}_I,\tens{P}_Q,\tens{P}_U,\tens{P}_V \} ,$$
the inhomogeneous part plus the identity. We will include the 
dilatation transformation in the semi-simple algebra set for easiness.
If we write the transfer equation as
$$\frac{d}{d\tau}\vec{I}=\tens{H}\vec{I},$$
then, $\tens{H}$, an element of the Poincar{\'e} group plus dilatations, can be 
decomposed into
$$\tens{H} = -\tens{K} + \tens{P},$$
where $\tens{K}$ is the usual absorption matrix, an element of the 
homogeneous group, and $\tens{P}$, which stands for the set of four translations introduced in Section 3, is an element of both the  radical and the 
inhomogeneous part of the equation.
In the last section we dealt with the homogeneous equation and found a 
solution for the evolution operator $\tens{O} (\tau ,\tau_0)$
by using the Lorentz group plus dilatations. 
Now, it is easy to demonstrate that if we are able to solve the equation
$$\frac{d}{d t}\vec{I}_R=\left( \tens{O}^{-1}(t,\tau _0)\tens{P}(t)
\tens{O}(t,\tau _0)\right)\vec{I}_R,$$
a solution for the complete transfer equation can be written in the form
\begin{equation}
\vec{I}(\tau)=\tens{O}(\tau ,\tau _0) \vec{I}_R (\tau ).
\label{sol}
\end{equation}
In fact, this result is exactly equivalent to the formal solution given
by Landi  Degl'Innocenti \& Landi  Degl'Innocenti(1985). To prove it, we note 
that $\tens{O}(t,\tau _0) \vec{I}_R$ will give, by properties of the
evolution operator, a new $\vec{I}_R(t)$. The effect of $\tens{P}$ is 
however independent of the actual value of $\vec{I}_R(t)$, we will always 
obtain that
$$ \tens{P} \vec{I}_R(t) =\pmatrix {j_I \cr j_Q \cr j_U \cr j_V}=\vec{J},$$
where the $j_i$'s are the infinitesimal parameters of the translation 
transformation: the emission vector in our particular case. The previous 
equation results therefore in
$$\frac{d}{d t}\vec{I}_R=\tens{O}^{-1}(t,\tau _0)\vec{J}(t),$$
which can be integrated at once:
$$\vec{I}_R (\tau)=\vec{I}_R (\tau_0)+\int_{\tau_0}^{\tau}\tens{O}^{-1}
(t,\tau _0)\vec{J}(t)dt.$$
Combining it with the homogeneous solution, we obtain the final complete solution:
$$\vec{I}(\tau)=\tens{O}(\tau,\tau _0)\left(\vec{I}_R (\tau_0) +\int_{\tau_0}^{\tau}\tens{O}^{-1}(t,\tau _0)\vec{J}(t)dt\right).$$
And benefiting from the well known properties of the evolution operator, we 
can transform this expression into the formal solution given in the above 
referred  paper:
$$ \vec{I}(\tau)=\tens{O}(\tau,\tau _0) \vec{I}_R (\tau_0) + 
\int_{\tau_0}^{\tau}\tens{O}(\tau,t)\vec{J}(t)dt.$$

Hence, once the evolution operator is solved as shown in the previous section 
we can 
use this expression to obtain the complete solution. Instead of doing that, 
we shall proceed with
the techniques provided by the group theory and obtain a completely
equivalent but independent expression for $\vec{I}_R$.

$\tens{P}$ belongs to the radical which, by definition, is an ideal of the 
Poincar{\'e} group, so that in fact
the term $ \tens{O}^{-1}\tens{P}\tens{O}$ is just a linear
combination of the infinitesimal generators of $R$:
$$ \left( \tens{O}^{-1}\tens{P}\tens{O}\right) = \eta _I \bbbone +
D_0 \tens{P}_I + D_1\tens{P}_Q + D_2\tens{P}_U+D_3\tens{P}_V .$$
Obtaining the coefficients $D_i$ ( with $i=0,1,2,3$) is 
quite long, the detailed calculation  is to be found in the Appendix. 
This calculation constitutes by 
itself a  demonstration of the first statement of this paragraph for our 
particular case, a long one, but which 
does not require further knowledge in group theory. The next step is to solve
the equation
$$\frac{d}{d\tau}\vec{I}_R= \left(D_0 \tens{P}_I + D_1\tens{P}_Q + 
D_2\tens{P}_U+D_3\tens{P}_V \right) \vec{I}_R .$$
This is in fact a very easy equation, as every $ \tens{P}_i$ commutes with 
each other. The solution can be given at once as
$$\vec{I}_R (\tau)= \mbox{e}^ { \tens{P}_I \int_{\tau_0}^{\tau} D_0(t)dt}
\cdot \mbox{e}^ { \tens{P}_Q \int_{\tau_0}^{\tau} D_1(t)dt}\cdot $$ 
$$\mbox{e}^ { \tens{P}_U \int_{\tau_0}^{\tau} D_2(t)dt}
\cdot \mbox{e}^ { \tens{P}_V \int_{\tau_0}^{\tau} D_3(t)dt} 
\vec{I}_R (\tau _0).$$
Calculation of the 
exponentials is straightforward: The $ \tens{P}_i$ are the infinitesimal 
generators of translations 
in the four axes $I,Q,U,V$, hence by exponentiation we recuperate the finite 
transformation:
\begin{equation}
 \vec{I}_R(\tau) =  \vec{I}_R (\tau _0) + 
\pmatrix {\int_{\tau_0}^{\tau} D_0(t)dt\cr \int_{\tau_0}^{\tau}  D_1(t)dt\cr 
\int_{\tau_0}^{\tau} D_2(t)dt\cr \int_{\tau_0}^{\tau} D_3(t)dt} = 
\vec{I}_R (\tau _0) + \vec{D}(\tau,\tau_0).
\end{equation}

The last step is to put together the homogeneous and inhomogeneous solutions
by using expression (\ref{sol}). We obtain
\begin{eqnarray}
\vec{I}(\tau)= \mbox{e}^{-\int_{\tau_0}^{\tau} \eta _I(t)dt}
\left[ \bbbone + g_3(\tau) \tens{H}_3 \right] \cdot 
\left[ \bbbone + g_2(\tau) \tens{H}_2 \right]\cdot \nonumber \\
\left[\cos g_1(\tau) \bbbone + \sin g_1(\tau)\tens{H_1} \right]
 \cdot\left[ \bbbone + g_6(\tau) \tens{L}_3 \right]
\cdot \left[ \bbbone + g_5(\tau) \tens{L}_2 \right]\cdot  \nonumber \\
\left[ \cos g_4(\tau) \bbbone + \sin g_4(\tau) \tens{L_1} \right]
\cdot  \left( \vec{I}(\tau _0) + \vec{D}(\tau,\tau_0) \right) .
\label{total}
\end{eqnarray}
Note that this solution is general for all the radiative transfer problems
known to date in polarization, provided the source function is given. As 
discussed in the previous section, it is even too general. 

This solution  is also independent of any model atmosphere. This is necessary
to ensure its generality, but presents the problem of the integrability: have
the integrals for the $g_i$'s and the inhomogeneous vector $\vec{D}$ 
an analytical expression for all and every interesting case? The most likely
 answer
is no. But whatever the answer, physical intuition indicates that there must 
always  exist at least a numerical solution  to them. However further work 
must be developed on the subject.

It is also important to note
that we are proposing not just an expression as solution of the RTE, but a
method: particular cases may ask for different orderings of the
generators or even a different decomposition of the product of exponentials.
We have seen that we can in any case give a solution in the form of seven 
exponentials, but, for instance, when solving constant matrix atmospheres it
may be more interesting to consider only the product of 3 exponentials, one 
with the $\tens{H}_i$ generators a second one for the  $\tens{L}_i$'s, and a
last for the dilatations, or even a sole one, in which case the solution
for the evolution operator can be written at once:
$$ \tens{O}(\tau,\tau_0)=\exp -\tens{K}(\tau - \tau_0),$$
in accordance with Magnus' solution (Magnus, 1954) or with the scalar--like
exponential solution (Semel \& L{\'o}pez Ariste, 1999).
 For any number of exponentials, the
method will work, the sole problem being to solve the subsequent scalar linear
equations and integrals. In all the cases, a {\em compact and finite} 
expression for the solution is obtained and the problem is reduced to the 
ability to integrate scalar expressions.

\section{Discussion and conclusion}

In this paper we have introduced a new formalism to handle Stokes parameters
and radiative transfer equations for polarized light. In this formalism, 
the Stokes parameters appear as a 4--vector in a Minkowski--like
4--dimensional space, and its evolution in time  looks
mathematically as typical rotations, contractions  and 
translations in this space. These movements are completely described by
the transformations of the group of Poincar{\'e} plus dilatations, a $10+1$ 
dimension group, well-known 
from other areas of physics and mathematics. The RTE is shown to be an
infinitesimal transformation of this group. We therefore propose that  
a solution to the RTE
can be given in the form of a finite transformation of the Poincar{\'e} plus 
dilatations group. Obtaining of this
solution from the variables present in the transfer equation raises some 
technical difficulties which have been overcome by the use of the Wei-Norman
method \cite{WN63}. The final obstacle is reduced to a scalar Riccati equation.

The Riccati equation is a well studied first order differential equation, 
characterized by its quadratic term. This non-linearity can at worst prevent
an explicit solution, and  usually  make it difficult to calculate. In any 
case the 
problem of giving a solution for the RTE will have been reduced from solving
a 4--dimensional vector equation to solving a scalar Riccati one. Whenever
this Riccati  equation can be integrated, a complete solution is
obtained for the RTE. 

Until now only numerical integration methods (see for instance Rees 
et. al, 1989, Bellot Rubio et. al, 1998 or
L{\'o}pez Ariste \& Semel, 1999) were capable of integrating non-constant
\nocite{RM89,BRRCC98,art2}
$\tens{K}$ matrices. The only way to test the validity of the solution and the
convergence rates was to compare them with previous methods, known
to converge asymptotically. The  solution   presented in this
paper may allow a comparison with an  analytically exact
solution. We anticipate that new numerical methods will be developed
taking advantage of the analytical solution; perhaps faster and more precise 
than previous ones. 

In order to obtain this solution we made use of  a mathematical frame, 
group theory, rarely seen in the astrophysical literature. The advantages 
gained
in the integration of the polarized RTE warranted the efforts. We anticipate
that new results in the
study of polarized light transfer in astrophysical problems will be achieved
by the use of this and related techniques.

\appendix
\section{Solution for the radical}

As explained in Section \ref{homo}, we need to calculate, in order to obtain
the final solution, a term of the form
\begin{equation}
\left( \tens{O}^{-1}\tens{P}\tens{O}\right),
\label{expr}
\end{equation}
where $\tens{O}$ is the evolution operator, solution of the homogeneous 
equation, i.e.,
$$ \tens{O}(\tau,\tau _0)=\mbox{e}^{g_3(\tau )\tens{H}_3}
\mbox{e}^{g_2(\tau )\tens{H}_2}\mbox{e}^{g_1(\tau )\tens{H}_1}
\mbox{e}^{g_6(\tau )\tens{L}_3}\mbox{e}^{g_5(\tau )\tens{L}_2}
\mbox{e}^{g_4(\tau )\tens{L}_1}\mbox{e}^{g_7 (\tau )\bbbone},$$
 and where $\tens{P}$ is the inhomogeneous part of the transfer equation
which can be written as
$$ \tens{P}= j_I(\tau ) \tens{P}_I+ j_Q(\tau ) \tens{P}_Q+
j_U (\tau )\tens{P}_U+ j_V(\tau ) \tens{P}_V ,$$
where the $j_i$ are the components of the emission vector. In a further 
effort to simplify the calculations, instead of this linear combination
we will use 
$$ \tens{P}= j_I\tens{P}_I+j_Q\tens{P}_Q+j_A\tens{P}_A+ j_B\tens{P}_B ,$$
where $\tens{P}_A=\tens{P}_U+i\tens{P}_V$, and $\tens{P}_B$ is 
its complex conjugate. Consequently, $j_A$ and $j_B$ are given by 
\begin{eqnarray*}
 j_A =\frac{1}{2}(j_U-ij_V) \\ j_B=\frac{1}{2}(j_U+ij_V ).
\end{eqnarray*}
Working out expression (\ref{expr}) implies the use of commutators of
$\tens{H_i}$, $\tens{L_i}$ and the dilatations with $\tens{P}_i$. Those 
commutators, which can be found in any textbook on group theory, are, for 
$\tens{P}_I$
\begin{eqnarray*}
\left[\tens{P}_I,\tens{H_1}\right]&=&-i\tens{P}_Q \\
\left[\tens{P}_I,\tens{H_2}\right]&=&-\tens{P}_B \\
\left[\tens{P}_I,\tens{H_3}\right]&=&-\tens{P}_A .\end{eqnarray*}
For $\tens{P}_Q$ we have
\begin{eqnarray*}
\left[\tens{P}_Q,\tens{H_1}\right]&=&-i\tens{P}_I\\
\left[\tens{P}_Q,\tens{H_2}\right]&=&\tens{P}_A \\
\left[\tens{P}_Q,\tens{H_3}\right]&=&-\tens{P}_A .\end{eqnarray*}
For $\tens{P}_A$
\begin{eqnarray*}
\left[\tens{P}_A,\tens{H_1}\right]&=&i \tens{P}_A \\
\left[\tens{P}_A,\tens{H_2}\right]&=&-2 \tens{P}_I \\
\left[\tens{P}_A,\tens{H_3}\right]&=& 0 .\end{eqnarray*}
And for $\tens{P}_B$
\begin{eqnarray*}
\left[\tens{P}_B,\tens{H_1}\right]&=&-i \tens{P}_B \\
\left[\tens{P}_B,\tens{H_2}\right]&=&-2 \tens{P}_Q \\
\left[\tens{P}_B,\tens{H_3}\right]&=&-2 (\tens{P}_I-\tens{P}_Q) .\end{eqnarray*}
Commutators for  the $\tens{L_i}$ can be obtained as the complex 
conjugated of the corresponding ones for the $\tens{H_i}$. Finally, for the
dilatation operator $\tens{D}$ (expressed in matrix representation by the 
identity matrix), we have:
\begin{equation}
[\tens{P}_i,\tens{D}]=\tens{P}_i,
\end{equation}
with $i=I,Q,A,B$.

Once we have all the rules of the game we can begin to play with expression
(\ref{expr}) and calculate its first term: 
$$R_1=\mbox{e}^{-g_3\tens{H}_3} \tens{P} \mbox{e}^{g_3\tens{H}_3}$$
(in what follows and for the sake of clarity we leave out the dependences on 
$\tau$
of $g_i$ and $j_i$ to recuperate them in the final expressions).
To this end we will need to calculate and add afterwards all the terms of 
the form
$$ \mbox{e}^{-g_3\tens{H}_3}\tens{P}_i\mbox{e}^{g_3\tens{H}_3}.$$ Each one
of which is to be calculated using an equivalent of the Baker--Hausdorff 
formula, which, for example for $\tens{P}_I$, affirms that
$$
 \mbox{e}^{-g_3\tens{H}_3}\tens{P}_I\mbox{e}^{g_3\tens{H}_3} = 
  \tens{P}_I + g_3 \left[\tens{P}_I,\tens{H}_3\right] + \frac{1}{2!}
g_3^2\left[ \left[\tens{P}_I,\tens{H}_3\right],\tens{H}_3\right]+ \ldots 
$$
The result of these calculations is
\begin{eqnarray*}
\mbox{e}^{-g_3\tens{H}_3}\tens{P}_I\mbox{e}^{g_3\tens{H}_3}&=&
\tens{P}_I-g_3\tens{P}_A ,\\
\mbox{e}^{-g_3\tens{H}_3}\tens{P}_Q\mbox{e}^{g_3\tens{H}_3}&=&
\tens{P}_Q-g_3\tens{P}_A ,\\
\mbox{e}^{-g_3\tens{H}_3}\tens{P}_A\mbox{e}^{g_3\tens{H}_3}&=&\tens{P}_A ,\\
\mbox{e}^{-g_3\tens{H}_3}\tens{P}_B\mbox{e}^{g_3\tens{H}_3}&=&
\tens{P}_B + 2g_3 (\tens{P}_I-\tens{P}_Q ).\end{eqnarray*}
So one obtains
\begin{eqnarray}
 R_1&=&(j_I+2g_3j_B)\tens{P}_I + 
(j_Q-2g_3j_B)\tens{P}_Q+ \nonumber \\ && (-j_I g_3-j_Q g_3+j_A)\tens{P}_A + 
j_B\tens{P}_B =\nonumber \\ &=&  c_{10}\tens{P}_I+c_{11}\tens{P}_Q+
c_{12}\tens{P}_A+c_{13}\tens{P}_B .\end{eqnarray}
The meaning of the coefficients $c_{1i}$ is self-evident. Next term is
$$R_2=\mbox{e}^{-g_2\tens{H}_2} R_1 \mbox{e}^{g_2\tens{H}_2} .$$
Partial results involved are
\begin{eqnarray*}
\mbox{e}^{-g_2\tens{H}_2}\tens{P}_I\mbox{e}^{g_2\tens{H}_2} =
S_0 \tens{P}_I + S_1\tens{P}_Q + S_2\tens{P}_A+S_3\tens{P}_B \\
\mbox{e}^{-g_2\tens{H}_2}\tens{P}_Q\mbox{e}^{g_2\tens{H}_2} =
S_0 \tens{P}_Q-S_1\tens{P}_I-S_2\tens{P}_A+S_3\tens{P}_B \\
\mbox{e}^{-g_2\tens{H}_2}\tens{P}_A\mbox{e}^{g_2\tens{H}_2} =
S_0 \tens{P}_A+S_1\tens{P}_B+2S_2\tens{P}_I-2S_3\tens{P}_Q \\
\mbox{e}^{-g_2\tens{H}_2}\tens{P}_B\mbox{e}^{g_2\tens{H}_2} =
S_0 \tens{P}_B+S_1\tens{P}_A+2S_2\tens{P}_Q-2S_3\tens{P}_I,\end{eqnarray*}
where the $S_0 , S_1 , S_2 , S_3$ are shortcuts for
\begin{eqnarray}
S_0 &=& \cosh g_2 \cdot \cos g_2 ,\\
S_1 &=& \sinh g_2 \cdot \sin g_2 ,\\
S_2 &=& -\frac{1}{2}(\cosh g_2 \cdot \sin g_2 + \cos g_2 \cdot \sinh g_2 ), \\
S_3 &=& \frac{1}{2}(\cosh g_2 \cdot \sin g_2 - \cos g_2 \cdot \sinh g_2 ). 
\end{eqnarray}
The result for $R_2$ is
\begin{eqnarray}
R_2&=&(c_{10}S_0-c_{11}S_1+2c_{12}S_2-2c_{13}S_3)\tens{P}_I + \nonumber \\&&
(c_{10}S_1+c_{11}S_0-2c_{12}S_3+2c_{13}S_2)\tens{P}_Q \nonumber \\& & +
(c_{10}S_2+c_{11}S_3+c_{12}S_0+c_{13}S_1)\tens{P}_A+ \nonumber \\& &
(c_{10}S_3-c_{11}S_2+c_{12}S_1+c_{13}S_0)\tens{P}_B \nonumber \\ &=&
c_{20}\tens{P}_I+c_{21}\tens{P}_Q+c_{22}\tens{P}_A+
c_{23}\tens{P}_B .
\end{eqnarray}
Next term is $$R_3=\mbox{e}^{-g_1\tens{H}_1} R_2 \mbox{e}^{g_1\tens{H}_1}$$
and, by means of the following partial results:
\begin{eqnarray*}
\mbox{e}^{-g_1\tens{H}_1}\tens{P}_I\mbox{e}^{g_1\tens{H}_1} &=&
\cos g_1 \tens{P}_I - i\sin g_1 \tens{P}_Q \\
\mbox{e}^{-g_1\tens{H}_1}\tens{P}_Q\mbox{e}^{g_1\tens{H}_1} &=&
\cos g_1 \tens{P}_Q - i\sin g_1 \tens{P}_I \\
\mbox{e}^{-g_1\tens{H}_1}\tens{P}_A\mbox{e}^{g_1\tens{H}_1} &=&
\mbox{e}^{ig_1}\tens{P}_A \\
\mbox{e}^{-g_1\tens{H}_1}\tens{P}_B\mbox{e}^{g_1\tens{H}_1} &=&
\mbox{e}^{- ig_1}\tens{P}_B   \end{eqnarray*}
one gets
\begin{eqnarray}
R_3&=&(c_{20}\cos g_1-ic_{21}\sin g_1)\tens{P}_I + \nonumber \\ &&+
(c_{21}\cos g_1-ic_{20}\sin g_1)\tens{P}_Q +\nonumber \\
&&+ c_{22} \mbox{e}^{ig_1}\tens{P}_A+
c_{23}\mbox{e}^{-ig_1}\tens{P}_B = \nonumber \\
&=& c_{30}\tens{P}_I+c_{31}\tens{P}_Q+c_{32}\tens{P}_A+c_{33}\tens{P}_B .
\end{eqnarray}
Now the process is to be repeated for $\tens{L}_i$ to obtain $R_4$, $R_5$
and $R_6$. Being the $\tens{L}_i$ the complex conjugated of $\tens{H}_i$, 
every expression is immediate just by using the corresponding complex 
conjugated coefficients and by substituting the functions $g_4$, $g_5$ and
$g_6$ for $g_1$, $g_2$ and $g_3$ respectively. We successively
obtain
\begin{eqnarray}
R_4&=&(c_{30}+2g_6c_{33})\tens{P}_I + 
(c_{31}-2g_6c_{33})\tens{P}_Q +\nonumber \\
&&+ (-c_{30}g_6-c_{31} g_6+c_{32})\tens{P}_A +
c_{33}\tens{P}_B \nonumber \\&=&  c_{40}\tens{P}_I+c_{41}\tens{P}_Q+
c_{42}\tens{P}_A+c_{43}\tens{P}_B ,\end{eqnarray}
and
\begin{eqnarray}
R_5=&=&(c_{40}T_0-c_{41}T_1+2c_{42}T_2-2c_{43}T_3)\tens{P}_I +\nonumber \\
&&+(c_{40}T_1+c_{41}T_0-2c_{42}T_3+2c_{43}T_2)\tens{P}_Q +\nonumber \\& & +
(c_{40}T_2+c_{41}T_3+c_{42}T_0+c_{43}T_1)\tens{P}_A+\nonumber \\&&+
(c_{40}T_3-c_{41}T_2+c_{42}T_1+c_{43}T_0)\tens{P}_B \nonumber \\ &=&
c_{50}\tens{P}_I+c_{51}\tens{P}_Q+c_{52}\tens{P}_A+
c_{53}\tens{P}_B ,
\end{eqnarray}
where
\begin{eqnarray}
T_0 &=& \cosh g_5 \cdot \cos g_5 ,\\
T_1 &=& \sinh g_5 \cdot \sin g_5 ,\\
T_2 &=& -\frac{1}{2}(\cosh g_5 \cdot \sin g_5 + \cos g_5 \cdot \sinh g_5 ), \\
T_3 &=& \frac{1}{2}(\cosh g_5 \cdot \sin g_5 - \cos g_5 \cdot \sinh g_5 ).
\end{eqnarray}
 The final result is
\begin{eqnarray}
R_6&=&(c_{50}\cos g_4+ic_{51}\sin g_4)\tens{P}_I +\nonumber \\ &&+
(c_{51}\cos g_4+ic_{50}\sin g_4)\tens{P}_Q+ c_{52} \mbox{e}^{-ig_4}\tens{P}_A+
\nonumber \\ &&+c_{53}\mbox{e}^{  ig_4}\tens{P}_B = \nonumber \\
&=&c_{60} \tens{P}_I + c_{61}\tens{P}_Q + c_{62}\tens{P}_A+c_{63}\tens{P}_B.
\end{eqnarray}

And we are only left with the dilatation operator, for which the operations
are at this point almost immediate and give:
\begin{equation}
\left( \tens{O}^{-1}\tens{P}\tens{O}\right) = \mbox{e}^{g_7}\left( c_{60} 
\tens{P}_I + c_{61}\tens{P}_Q + c_{62}\tens{P}_A+c_{63}\tens{P}_B\right).
\end{equation}
The $D_i$ coefficients at section 5, can straightforwardly be obtained from
this expression as
\begin{eqnarray}
D_0(\tau)=\mbox{e}^{g_7(\tau)}c_{60}(\tau) \nonumber \\ 
D_1(\tau)=\mbox{e}^{g_7(\tau)}c_{61}(\tau)   \nonumber \\
D_2(\tau) = \mbox{e}^{g_7(\tau)}\frac{1}{2}(c_{62}(\tau)+c_{63}(\tau)) \nonumber \\
D_3(\tau)= \mbox{e}^{g_7(\tau)}\frac{1}{2} i(c_{63}(\tau)-c_{62}(\tau)).
\end{eqnarray}

\begin{acknowledgements}
The authors are indebted to M. Landolfi and M. Landi Degl'Innocenti for 
precious discussions and comments.
\end{acknowledgements}

\bibliographystyle{aabib99}

\end{document}